\documentclass{vgtc}                          

\ifpdf
  \pdfoutput=1\relax                   
  \usepackage{graphicx}                
  \DeclareGraphicsExtensions{.pdf,.png,.jpg,.jpeg} 
\else
  \usepackage{graphicx}                
  \DeclareGraphicsExtensions{.eps}     
\fi%

\graphicspath{{figures/}{pictures/}{images/}{./}} 

\usepackage{enumitem}
\usepackage{microtype}                 
\PassOptionsToPackage{warn}{textcomp}  
\usepackage{textcomp}                  
\usepackage{mathptmx}                  
\usepackage{times}                     
\usepackage{cite}                      
\usepackage{tabu}                      
\usepackage{booktabs}                  
\usepackage{soul}
\usepackage{todonotes}
\usepackage{caption}
\usepackage{subcaption}
\usepackage{wrapfig}
\usepackage{amsmath}
\usepackage{hyperref}

\onlineid{0}

\vgtccategory{Research}

\title{Bayesian Modelling of Alluvial Diagram Complexity}

\author{Anjana Arunkumar, \textit{Student Member, IEEE}\thanks{e-mail: aarunku5@asu.edu} %
\and Shashank Ginjpalli\thanks{e-mail: sginjpal@asu.edu} %
\and Chris Bryan, \textit{Member, IEEE}\thanks{e-mail: chris.bryan@asu.edu}}
\affiliation{\scriptsize Arizona State University}

\teaser{
  \centering
  \includegraphics[width=.9\linewidth]{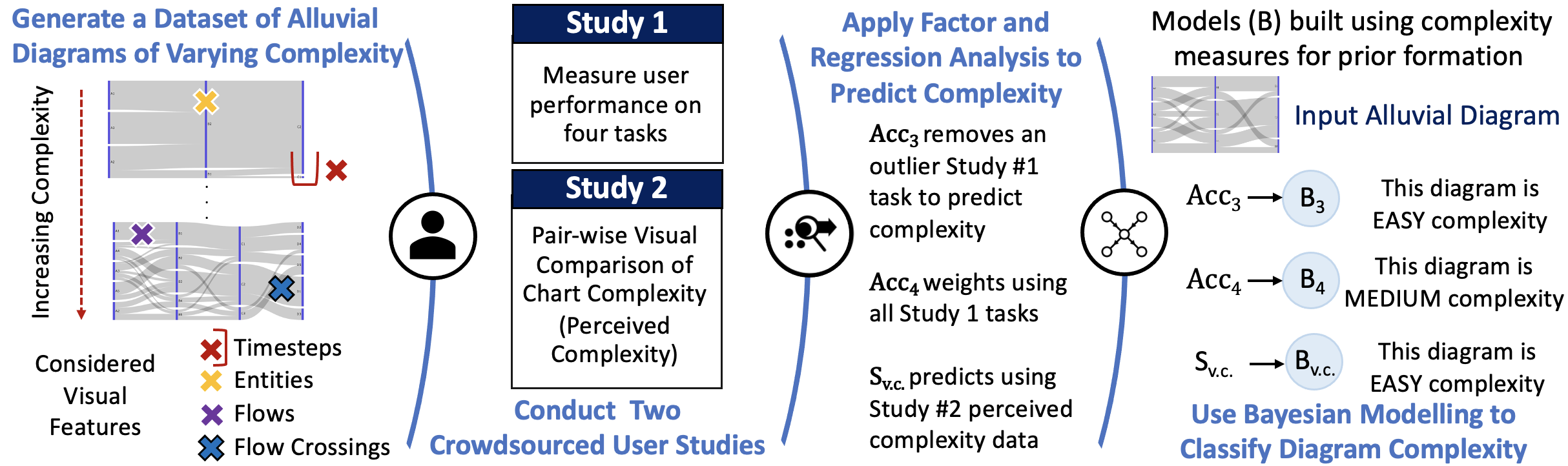}
  \caption{We study the complexity of alluvial diagrams with the following workflow: (1) We generate a publicly-available testing dataset of alluvial diagrams that vary complexity based on a combination of features. (2) We conduct two crowdsourced user studies to measure the task performance and perceived complexity of the charts. (3) Applying a combination of factor and regression analysis to weight feature contributions, (4) we model the complexity of alluvial diagrams.}
  \label{fig:teaser}
}

\abstract{Alluvial diagrams are a popular technique for visualizing flow and relational data. However, successfully reading and interpreting the data shown in an alluvial diagram is likely influenced by factors such as data volume, complexity, and chart layout. To understand how alluvial diagram consumption is impacted by its visual features, we conduct two crowdsourced user studies with a set of alluvial diagrams of varying complexity, and examine (i) participant performance on analysis tasks, and (ii) the perceived complexity of the charts. Using the study results, we employ Bayesian modelling to predict participant classification of diagram complexity. We find that, while multiple visual features are important in contributing to alluvial diagram complexity, interestingly the importance of features seems to depend on the type of complexity being modeled, i.e. task complexity vs. perceived complexity.

} 

\CCScatlist{
  \CCScatTwelve{Human-centered computing}{Visu\-al\-iza\-tion}{Visu\-al\-iza\-tion techniques}{Empirical studies in visualization}{};
}

\begin{document}

\maketitle

\section{Introduction}

\begin{wrapfigure}{r}{0.5\columnwidth}
    \centering
    \includegraphics[width=0.5\columnwidth]{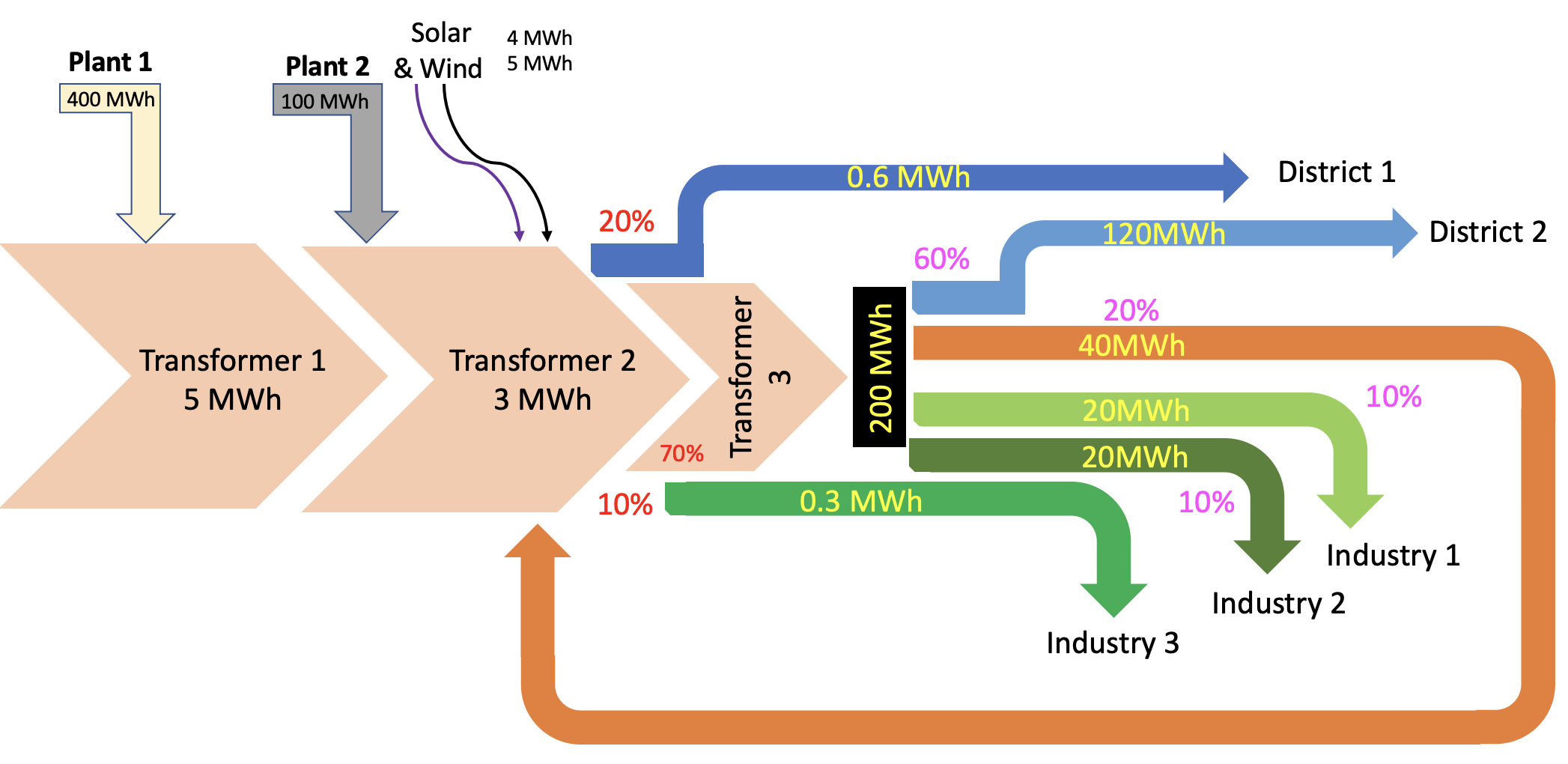}
    \vspace{-1em}
    \caption{An example Sankey diagram showing energy flow.}
    \label{fig:sankey}
\end{wrapfigure}

Alluvial diagrams are a type of Sankey diagram that visualize information flow between entity groups. In contrast to the general definition of Sankey diagrams (see Figure~\ref{fig:sankey}), which permit flow to go in any direction, alluvial diagrams group related entities into columns which are aligned along a common axis (e.g., going left-to-right), with the constraint that flow data can only belong to one entity in each group/column. As the height of the flow between entities shows data quantity, alluvial diagrams are a popular technique for visualizing time-varying network data, flow-based data, and relationships across multidimensional data, and have been applied in domains that include market/timeline analysis, network monitoring, and energy and power flows~\cite{1532152,ruan2017detecting,subramanyam2015using}. 

Unfortunately, interpreting an alluvial diagram can become a difficult task as the chart scales in size, complexity, and amount of information shown. Additionally, prior research has shown that more complex (and less familiar) visualizations, such as flow diagrams \cite{brambilla2012illustrative}, are less readable and therefore harder to interpret \cite{dimara2018task, dasgupta2017towards}. 

\setlength{\intextsep}{1.2em}%
\setlength{\columnsep}{1.2em}%

To better understand what makes alluvial diagrams ``complex'', we conduct and analyze two crowdsourced user studies. Figure~\ref{fig:teaser} shows the steps. (1)~We create a synthetic dataset of alluvial diagrams with varying estimated complexity (a value called $S_a$) by statistically controlling the underlying dataset properties for each chart. (2)~We conduct two crowdsourced user studies. In Study~\#1, participants perform four analysis tasks on the diagrams. In Study~\#2, participants compare pairs of diagrams to rate their relative, perceived complexities. 
(3)~Using the collected study data, we perform regression and factor analysis to weight the impact of visual features in determining complexity, (4) and use these as
Bayesian priors to classify alluvial diagrams for task performance and perceived complexity (i.e., label alluvial diagrams as having 
\textit{easy}, \textit{medium}, or \textit{hard} complexity). While we find that all of the considered visual features significantly contribute towards modelling complexity, we interestingly find that the most important factor changes depending on the type of complexity (i.e., task complexity vs. perceived complexity) being modeled.

To our knowledge, \textit{this is the first research that empirically assesses, quantifies, and models the complexity of alluvial diagrams}. Our modelling approach is also replicable for other types of visualization techniques. We additionally include a robust set of supplemental materials for use by the research community, including dataset generation scripts, chart datasets and rendered image files, and collected study data, publicly available at \url{https://tinyurl.com/386adhwf}.



\section{Related Work}

\textbf{Optimizing Alluvial Diagrams.}
As mentioned in the Introduction, alluvial diagrams are a specific type of Sankey diagram with additional constraints for showing flow and relational data. Similar to how node-link diagrams optimize the placement of nodes and edges for readability (such as via force-directed layouts), algorithms for creating alluvial diagrams also try to optimize the placement of entities/nodes
\cite{alemasoom2016energyviz,8365985}. For both techniques, a poor organization of entities/nodes can lead to increased edge/flow crossings, lowering the chart's readability. However, in contrast to node-link diagrams (and Sankey diagrams) which can freely place nodes anywhere, alluvial diagrams only allow (i) an entity to be moved within its column, and (ii) if the dataset is not time-based, columns may be swapped. Similar to minimizing edge crossings in node-link diagrams, minimizing flow crossings for alluvial diagrams is an NP-hard problem~\cite{garey1983crossing}. Recent optimization efforts for Sankey/alluvial diagrams have utilized linear programming~\cite{8365985, alemasoom2016energyviz}. The alluvial diagrams in our dataset are created using the Plotly library,\footnote{\url{https://plotly.com/python/}} which uses the d3-sankey package\footnote{\url{https://github.com/d3/d3-sankey}} for computing layout.

\textbf{Visualization Readability and Effectiveness.}
Broadly, the topic of graphical perception is concerned with how visualizations are perceived and interpreted~\cite{cleveland1987graphical}. Perception and readability depends on many factors, including the visual encodings being used (i.e., the marks and channels), the number of data dimensions being encoded \cite{iliinsky2011designing,munzner2014visualization,padilla2020powerful}, the amount of data shown \cite{keim2013big}, how the chart is styled \cite{kim2016data}, what rhetorical elements are present~\cite{hullman2011visualization}, and even the current cognitive focus of the person viewing the chart~\cite{healey2012attention}.

Bayesian statistics have been found to reliably model human cognition, and further allow for the principled incorporation of `irrational' behavior \cite{griffiths2010probabilistic,liu2008distributed,lieder2018empirical}. Bayesian modelling has been used to measure the change of people's beliefs on visualization viewing \cite{gelman2004exploratory,kim2019bayesian}, and has also been extrapolated to define a signal-detection approach to reason about visualization-based inference~\cite{hullman2019authors}. Such approaches can be extrapolated to predict what users find `interesting' in a visualization, and further to formalize the \textit{effectiveness} of visualization \textit{messaging}.

Accurate prior elicitation is a challenge in using Bayesian methods; to circumvent this, we directly set priors based on the visual features used as complexity control factors. Previous work has shown that frequency format based representation of information helps participants better contextualize priors~\cite{gigerenzer1995improve,wu2017towards}. We accordingly familiarize participants with the visual features of alluvial diagrams and how their counts are associated with diagram complexity during training.

Accuracy, task-driven evaluation, and perception of visual properties (such as color, shape, and size) in visualizations  \cite{zhu2007measuring,keim2010mastering,albers2014task,szafir2017modeling,smart2019measuring} have also been used to inform effective visualization design.  The framing of effectiveness in past work is closely related with our framing of complexity; however, to the best of our knowledge, there has been no definition of a mathematical formula for complexity for a given visualization type.

\section{Studies}

\textbf{Dataset.} To conduct our studies, we first created a dataset of synthetic alluvial diagrams. Each diagram is rendered as a $1920 \times 1080$ PNG image with a uniform styling of grey flow arcs, blue entity rectangles, and flow going left-to-right. 

A Python script created the underlying data for each chart by controlling the following four factors: (1)~The \textit{number of timesteps}, $t$, was randomly selected between 3--6. (2) The \textit{number of entities} for each timestep was randomly selected between 2--5. (3) The \textit{total flow} was either 30, 50, or 80 units, and was based on the amount of selected timesteps and entities. (4) Finally, the \textit{flow size} for each flow arc was between 25--50\% of an entity's total flow, with the constraint that the largest flow going into an entity was at least 5\% larger than the next largest flow for that entity. All bounds were set to ensure that flows in the generated diagrams were at least 10px thick, to make them sufficiently distinguishable for the purpose of study tasks. The resultant datasets were rendered using the Plotly Python library. For each created chart, we also record the \textit{number of flow crossings}, where two flow arcs cross over each other.

\begin{wrapfigure}{r}{0.5\columnwidth}
    \centering
    \includegraphics[width=0.5\columnwidth]{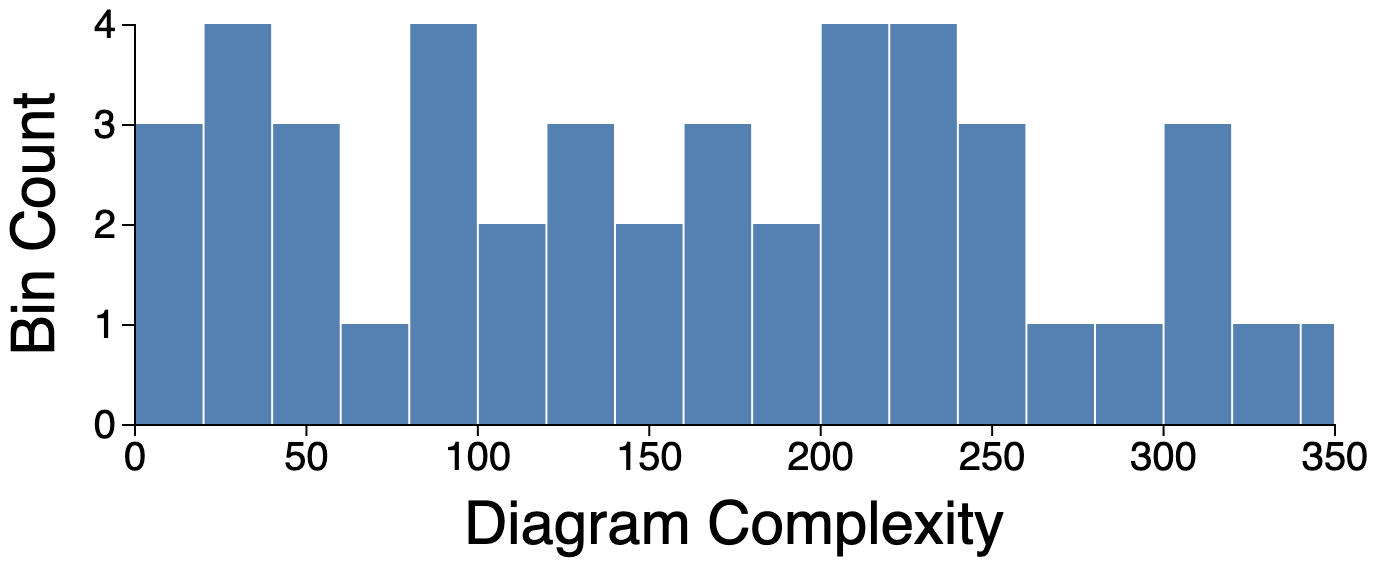}
    \vspace{-1em}
    \caption{This histogram shows the distribution of chart complexities of the alluvial diagrams used in the studies.}
    \label{fig:dataset_distribution}
\end{wrapfigure}

In total, we created $48$ charts, using Equation~1 to provide an \textit{estimated complexity score} $S_a$ for a given chart $a$ as the sum of its timesteps $t$, entities $e$, flow arcs $f$, and flow crossings $c$.\footnote{\textit{Total flow} is not considered in Equation~1 as chart sizes were normalized to fit the $1920 \times 1080$ PNG filesize.}

\begin{equation}
 S_a = t_a + e_a + f_a + c_a
 \label{eq:1}
\end{equation}
 
\begin{figure}[h]
    \centering
     \begin{subfigure}[b]{0.3\columnwidth}
         \centering
         \includegraphics[width=\textwidth]{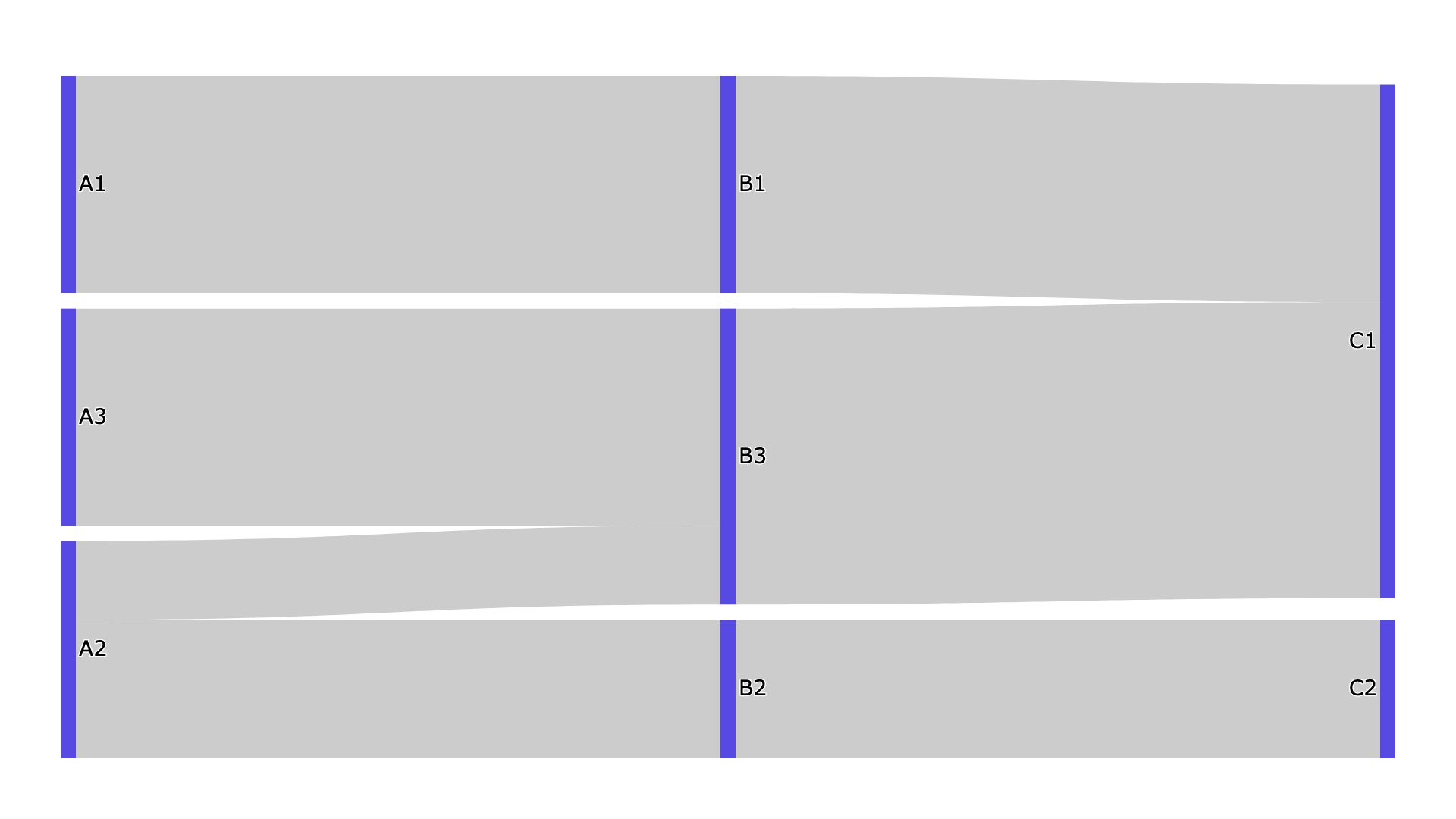}
         \caption{$S_1=17$}
         \label{fig:y equals x}
     \end{subfigure}
     \hfill
     \begin{subfigure}[b]{0.3\columnwidth}
         \centering
         \includegraphics[width=\textwidth]{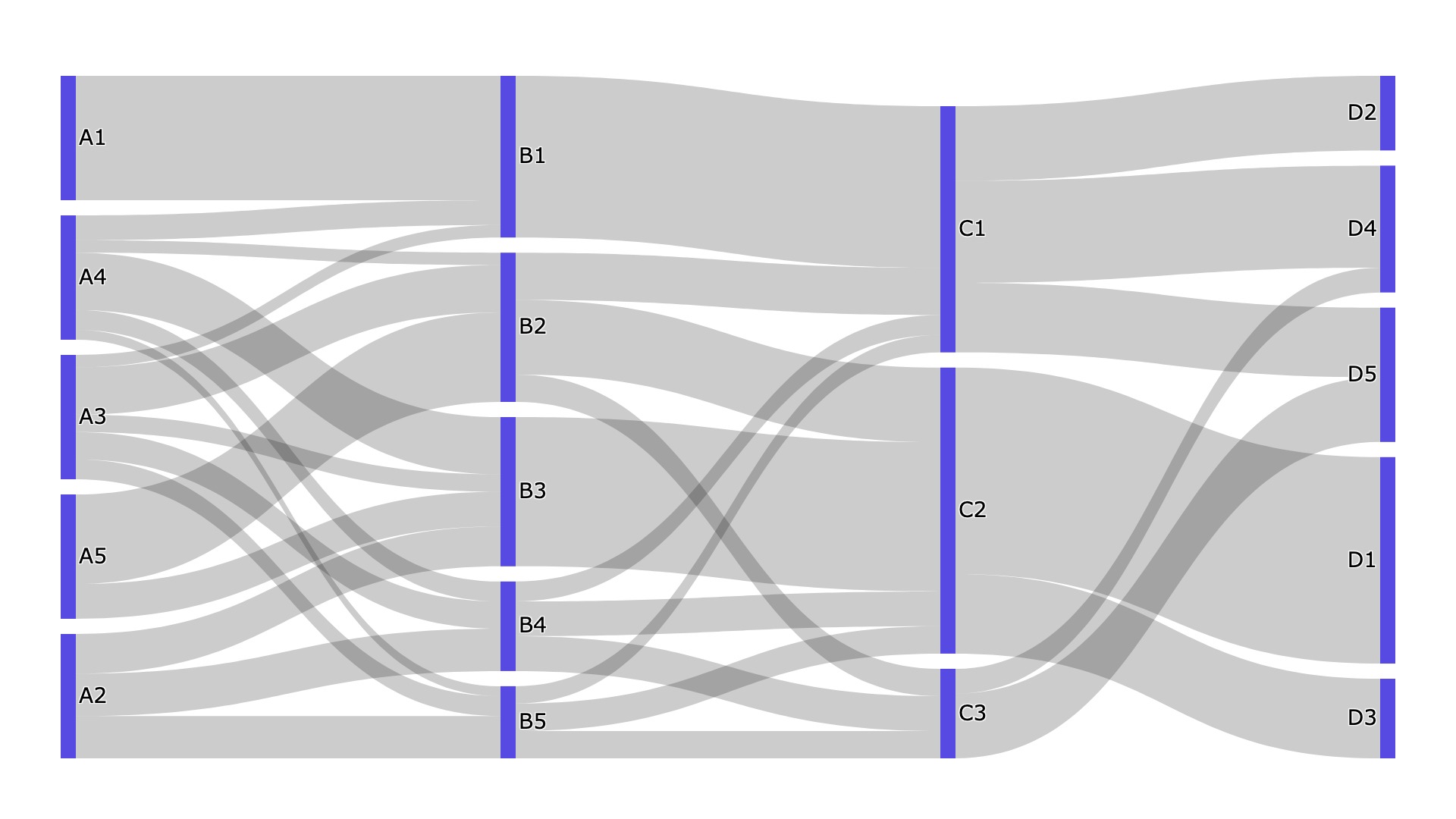}
         \caption{$S_{47}=204$}
         \label{fig:three sin x}
     \end{subfigure}
     \hfill
     \begin{subfigure}[b]{0.3\columnwidth}
         \centering
         \includegraphics[width=\textwidth]{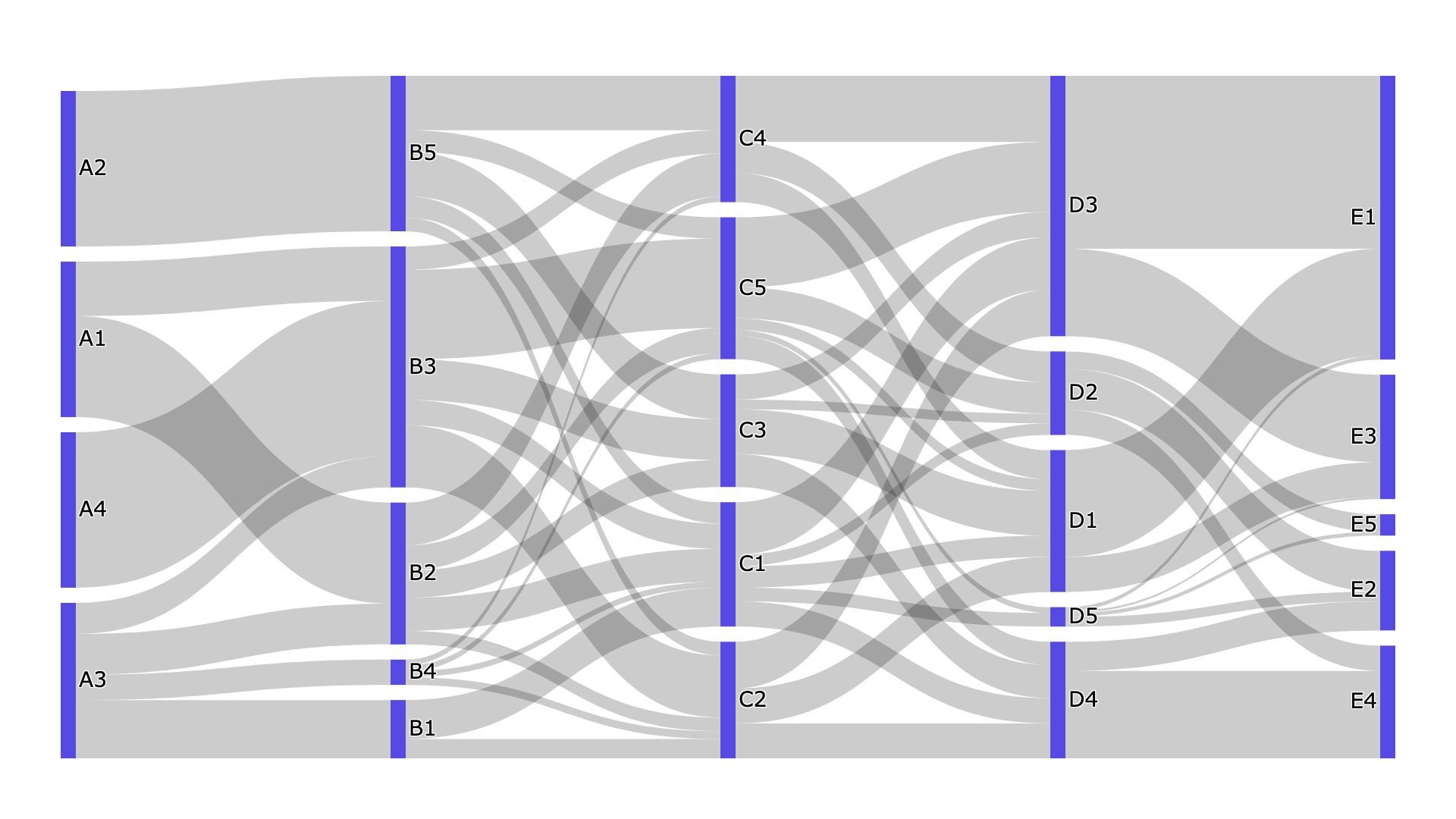}
         \caption{$S_{45}=306$}
         \label{fig:five over x}
     \end{subfigure}
        \caption{Three example alluvial diagrams used in the studies ($a=1, 47$, $45$), along with their estimated complexity score $S_a$.}
        \label{fig:example_alluvials}
\end{figure}

Figure~\ref{fig:dataset_distribution} shows the distribution of the 48 charts based on their $S_a$ values, and Figure~\ref{fig:example_alluvials} shows three examples of created charts. The full set of chart images and datasets (along with the Python script) is available in the supplementary materials. 

\textbf{Study~\#1 Design.}
Participants were shown a sequence of alluvial diagrams (one per trial) and asked to perform one of four tasks for each chart: (T1) Between which two timesteps are the most number of flows? (T2) Which entity in the diagram is the largest? (T3) Which flow arc in the diagram is the largest? (T4) Which entity has the most total flow activity (number of arcs entering and exiting)?

Each task requires a participant to examine the entire alluvial diagram to answer, entering their response into an input box. Of the 48 charts, 3 were reserved for training (each task was performed twice in training, for 8 total training trials). The remaining 45 charts were used for the main study. Each participant completed 19 trials during a session. For each trial, one chart and task were randomly selected and shown to the participant (no chart appeared twice during a session). An attention check appeared after the 10th question. Before running the main study, a pilot study was conducted with 3 participants to validate the design.

\textbf{Study~\#1 Participants.} We recruited 100 participants on Prolific\footnote{https://www.prolific.co} using the following user filters: (i) self-reported normal or corrected-to-normal vision, (ii) a first language of English, (iii) located in a U.S. college pursuing an undergraduate degree or higher, (iv) at least 100 previously completed Prolific surveys, and (v) an approval rate of 90\% or more.

Participants were paid $\$$2.50 for participation. Study~\#1 duration averaged just over 18.5 minutes (median 15:09), resulting in an average hourly pay of $\$$8.10 (median $\$$9.90). Qualtrics~\cite{ginn2018qualtrics} was used to display the study and store participant responses. Nine participants were excluded as their answers did not meet the expected format, or their performance indicated they did not understand the study questions during/after training.

\textbf{Study~\#2 Design.} For each trial in Study~\#2, participants were shown a pair of alluvial diagrams displayed side-by-side and asked to rate which chart was more complex (or if the charts were of approximately equal complexity).
Each participant completed 31 trials, where each trial randomly selected two of the 45 charts for comparison (990 total pairs possible). An attention check appeared after the 15th question. Three training trials were performed before beginning the main study. Similar to Study~\#1, a pilot study was first conducted to validate the design with 3 users.

\textbf{Study~\#2 Participants.} We recruited 150 participants on Prolific with the same filter settings as Study~\#1, again using Qualtrics to display trials and record responses. Participants were paid $\$$1.25 for participation. Study~\#2 duration was just over 7.5 minutes on average (median 6:06), resulting in an average hourly pay of $\$$10.00 (median $\$$12.29). No participants were excluded from this study, and each chart pair was rated either 4 or 5 times.

\begin{figure}[h]
    \centering
     \begin{subfigure}[b]{0.48\columnwidth}
         \centering
         \includegraphics[width=\textwidth]{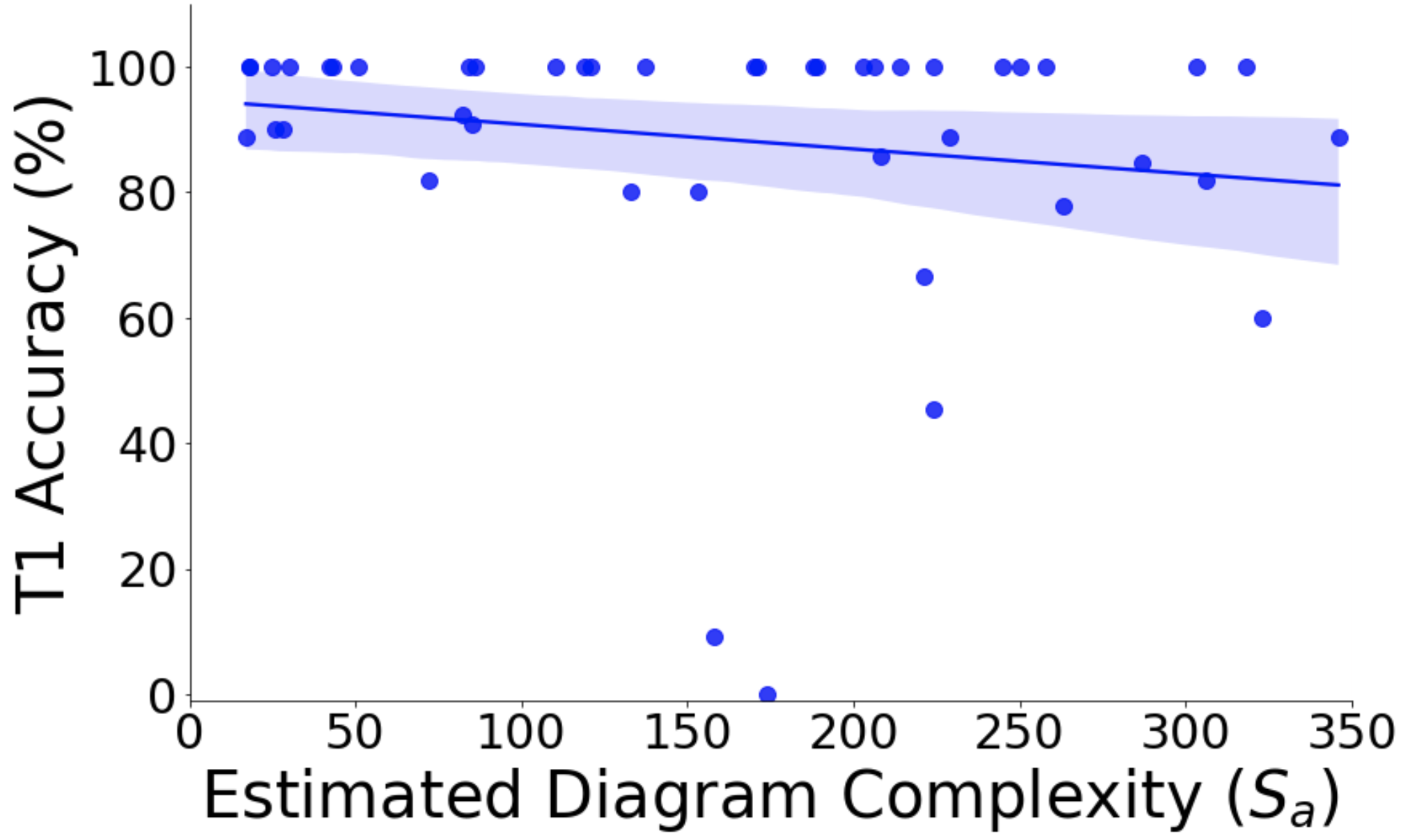}
         \caption{T1: Most Active Timestep, $R^2=0.656$}
         \label{fig:y equals x}
     \end{subfigure}
     \hfill
     \begin{subfigure}[b]{0.48\columnwidth}
         \centering
         \includegraphics[width=\textwidth]{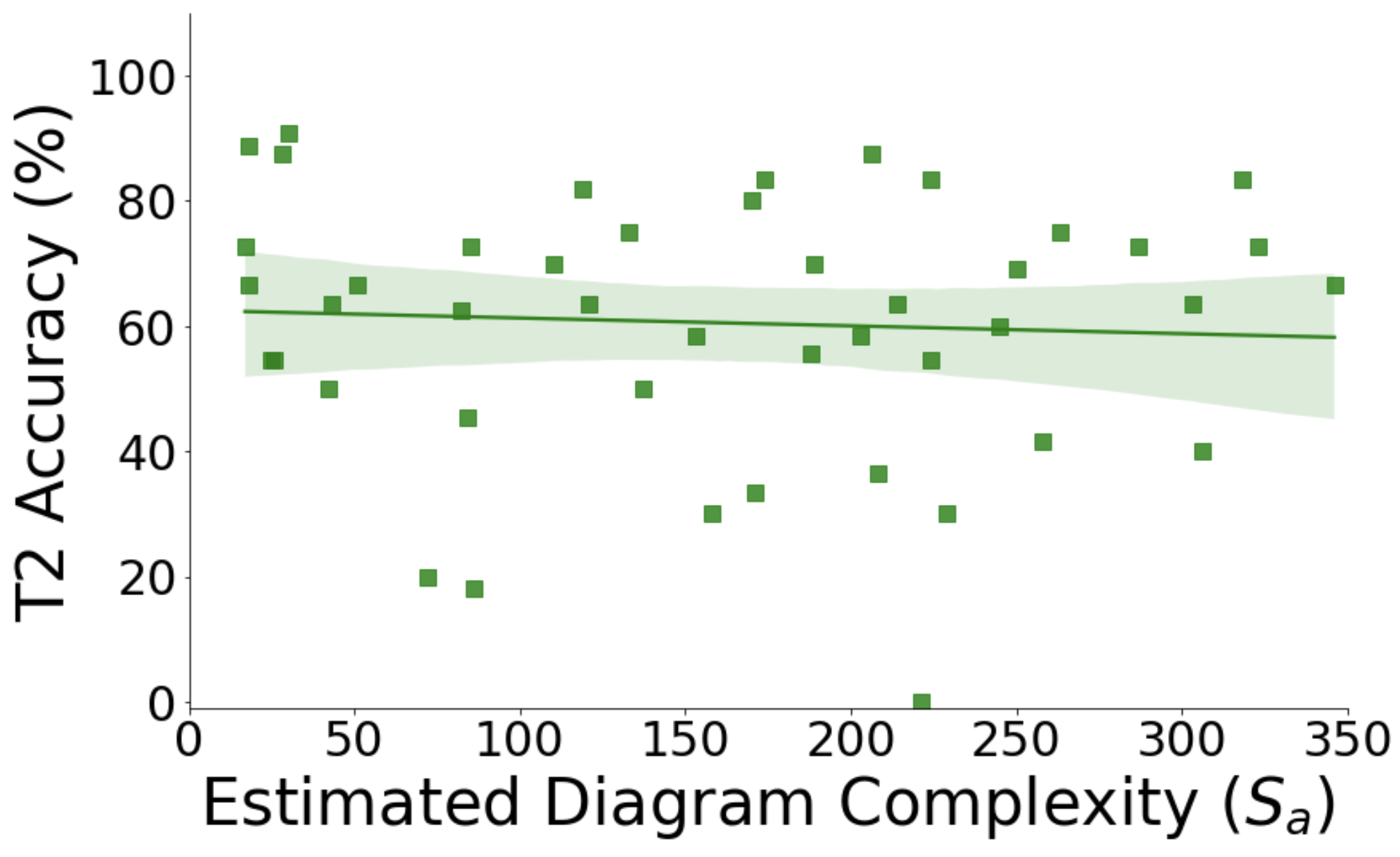}
         \caption{T2: Largest Entity, $R^2=0.642$}
         \label{fig:three sin x}
     \end{subfigure}
     \hfill
     \begin{subfigure}[b]{0.48\columnwidth}
         \centering
         \includegraphics[width=\textwidth]{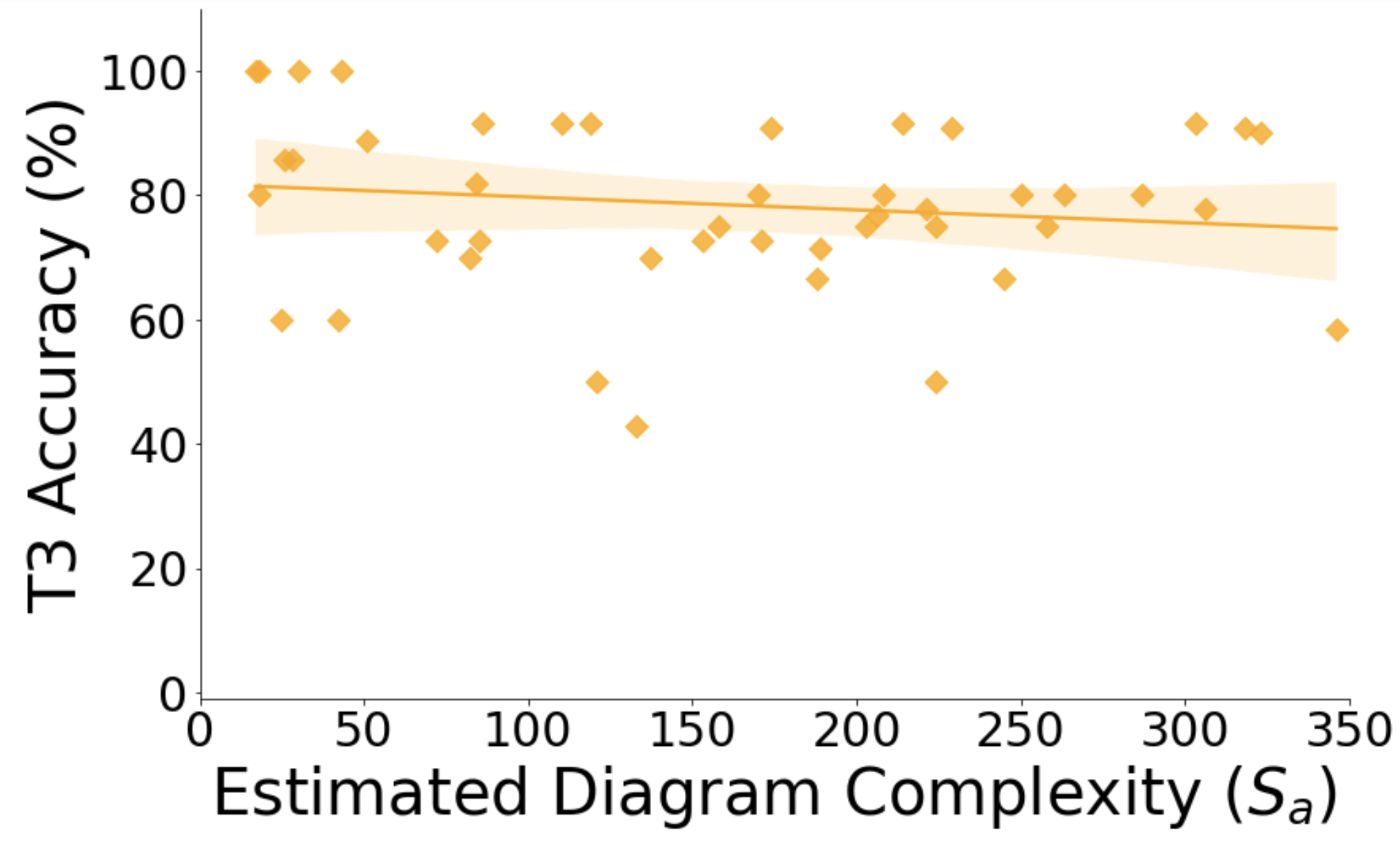}
         \caption{T3: Largest Flow, $R^2=0.690$}
         \label{fig:five over x}
     \end{subfigure}
     \hfill
     \begin{subfigure}[b]{0.48\columnwidth}
         \centering
         \includegraphics[width=\textwidth]{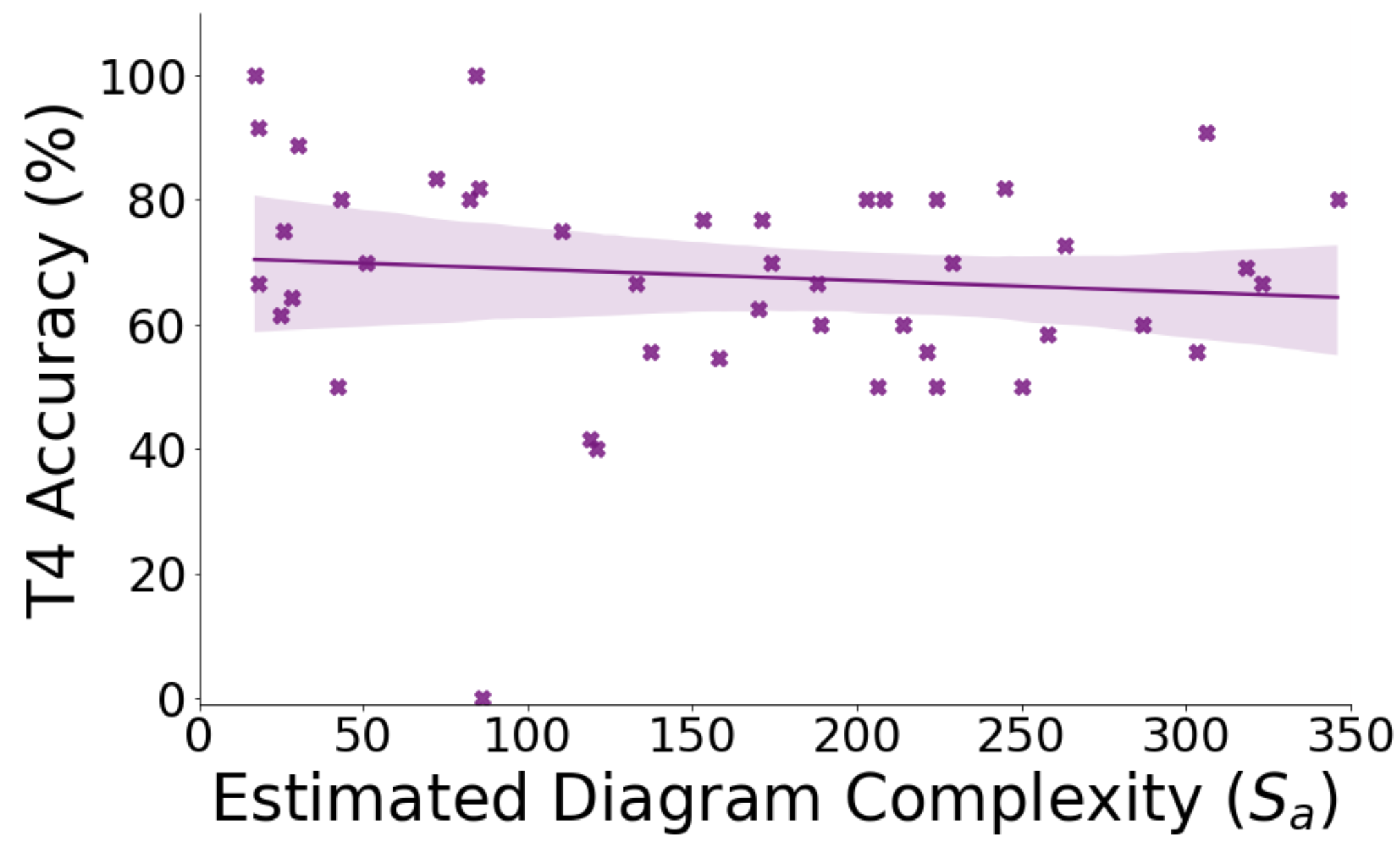}
         \caption{T4: Most Active Entity, $R^2=0.665$}
         \label{fig:five over x}
     \end{subfigure}
        \caption{Study~\#1 results: For tasks T1--T4, each point shows one alluvial diagram in the dataset, plotted by its task performance accuracy (vertical axis) and $S_a$ value (horizontal axis), fitted with a regression line and 95\% confidence interval bands. The $R^2$ value is also reported for each task ($p<0.001$ for all tasks).}
        \label{fig:study1_results}
\end{figure}

\vspace{-1em}
\begin{table}[h]
    \footnotesize
    \centering
    \begin{tabular}{lll}
        \toprule
        \textbf{Task} & \textbf{Accuracy (and SD)}\\\midrule
        T1: Most Active Timestep & $88.37 \pm 0.95$ \\
        T2: Largest Entity       & $57.10 \pm 2.04$\\
        T3: Largest Flow         & $79.40 \pm 1.59$\\
        T4: Most Active Entity & $68.93 \pm 2.00$\\
    \bottomrule
    \end{tabular}
    \vspace{-.5em}
    \caption{Performance results for Study 1.}
    \label{tab:1}
\end{table}

\section{Results}

\textbf{Study 1 Results.} Table~\#1 summarizes the task performance for T1--T4 (i.e., the percent of responses that were correct). T1 had (by far) the highest performance, and T2 the lowest. These results are reflected in Figure~\ref{fig:study1_results}, which plots, for each task, chart performance against its $S_a$ value. While the performance scores for individual chart-task pairs can vary, each of the four fitted regression lines indicates a similar trend: as a chart's $S_a$ value increases, its task performance decreases.



\begin{wrapfigure}{r}{0.5\columnwidth}
    \centering
    \includegraphics[width=0.48\columnwidth]{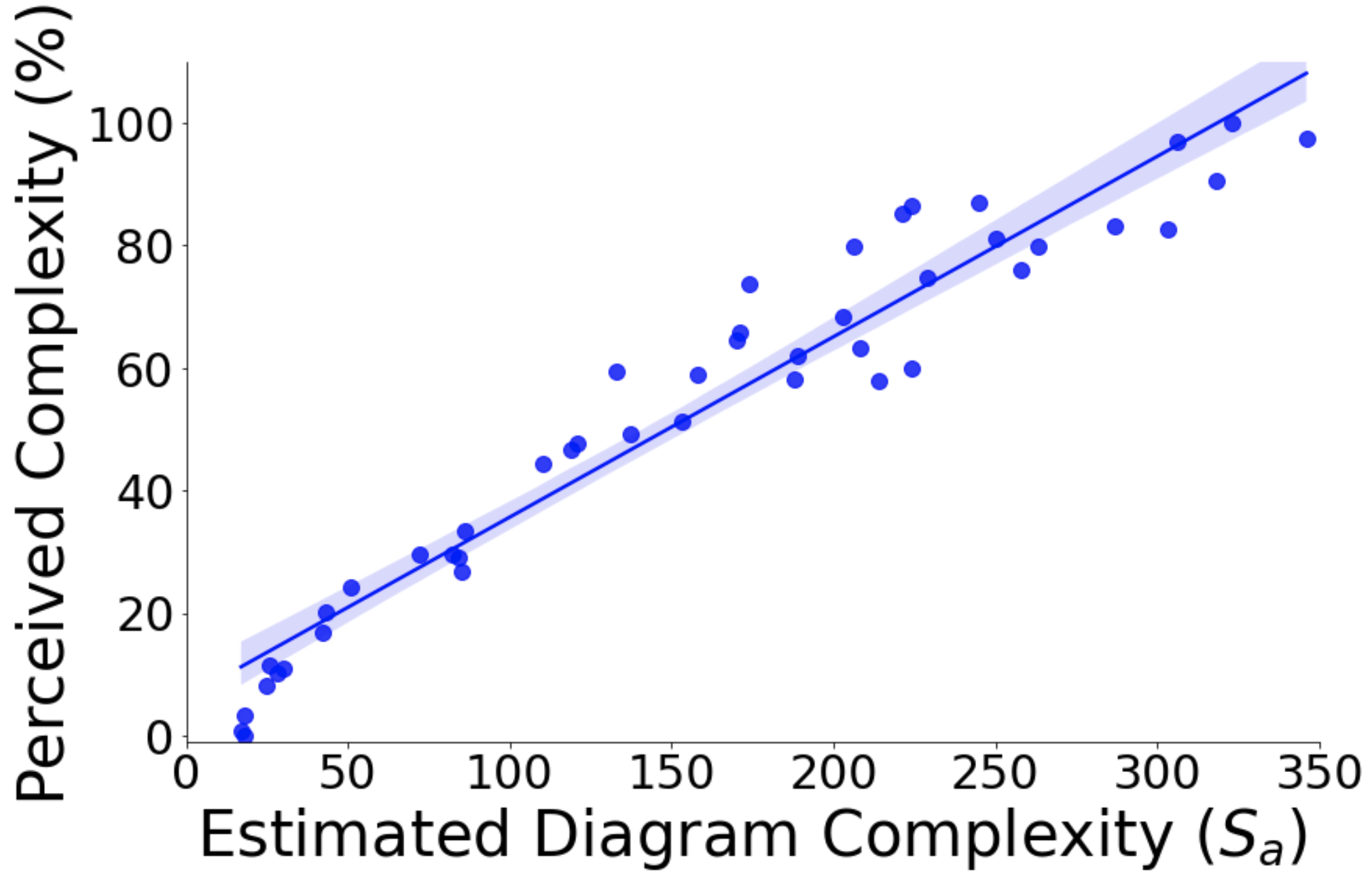}
    \vspace{-0.75em}
    \caption{Study~\#2 results: each point shows one alluvial diagram in the dataset plotted by its \textit{perceived complexity} (vertical axis) and its $S_a$ value (horizontal axis), fitted with a regression line ($R^2$=0.981, $p<0.001$) and 95\% confidence interval bands.}
    \vspace{-1.2em}
    \label{fig:study2Plot}
\end{wrapfigure}

\textbf{Study 2 Results.}
For each trial in Study~\#2, participants compared a pair of charts with relative complexity ratings. We transform these relative ratings into an overall list of complexity rankings with the following process: Each diagram starts with a score of 0, and $\pm$10 points are added based whether it is rated more/less complex in a trial. No points are added if two charts are rated equally complex in a trial. Summing the scores from all the trials determines a chart's overall \textit{perceived complexity}. Figure~\ref{fig:study2Plot} shows these overall \textit{perceived complexity} values plotted against $S_a$ values. Like Study~\#1, we fit a regression line which indicates a high correlation between a chart's \textit{perceived complexity} and its $S_a$ value.

\section{Bayesian Complexity Model}

Bayesian modelling has been used to measure the \textit{effectiveness} of visualizations~\cite{hullman2019authors}. Additionally, the visual features analyzed are frequency-based, and can therefore better inform participants of the significance of the prior, in line with previous work~\cite{wu2017towards}. We hence aim to mimic our study user's assessment of alluvial diagram complexity, except we bin the diagrams into discrete categories of \textit{easy, medium,} and \textit{hard} complexity, given the occurrence of considered visual features.

For each study, we analyze the effects of four independent variables, the visual features \textit{number of timesteps, number of flows, number of entities}, and \textit{number of flow crossings}, on the measured complexity data obtained from both studies (for Study~\#1, the performance for each task, and for Study~\#2, the perceived complexity scores). To fine-tune our estimated complexity score, we perform iterations of factor (PCA/Unrotated/Varimax) and linear/multiple regression analyses~\cite{hair2009multivariate,kefi2010using} to determine the effect strength of each factor (i.e., each visual feature); this is done with the aim of preserving statistical power during analyses.\footnote{Detailed statistics are fully reported in the supplementary material \label{f4}}

\textbf{Study 1 Modelling.} As a first step, we ran regressions of each visual feature against each task individually, to examine the nature and extent of their relationships\textsuperscript{\ref{f4}}. For each task, each of the four independent variables had a significant positive correlation (all p$<$0.001), though for all four tasks \textit{number of entities} had the highest $R^2$ value (T1=0.66, T2=0.641, T3=0.720, T4=0.682), which indicates it had the most impact on performance.

We next perform a factor analysis on the tasks, to verify if they are suitable for constructing summative dependent variables. We find that T2, T3, and T4 load onto a single factor, where T2 is strongly significant (-0.51), T3 (-0.37) and T4(-0.31) are weakly significant, and T1 (-0.22) fails to load. Accordingly, we develop a $Acc_3$, a summated dependent variable comprised of T2, T3, and T4; we also construct an alternate dependent variable, $Acc_4$, summated over all four tasks, to examine how $T1$'s inclusion impacts the importance of the four visual features.  We perform feature-wise regression against $Acc_3$ and $Acc_4$; \textit{number of entities} again serves as the strongest predictor ($R^2$ is 0.704 for $Acc_4$, 0.712 for $Acc_3$). 

On performing factor analysis over the four independent variables, we find all four features load onto a single component, and are all strongly significant (\textit{number of timesteps}=0.9, \textit{number of flows}=0.939, \textit{number of flow crossings}=0.86, \textit{number of entities}=0.930), where \textit{number of flows} has the highest significance. We construct an independent summated feature ($F$) variable, and run it against all the dependent variables (i.e., task performance); \textit{number of entities} remains the highest predictor throughout.

Next, we perform k-fold cross validation (k=5) to update Equation~\ref{eq:1} based on the standardized regression coefficients obtained for multiple regression of features against $Acc_3$ and $Acc_4$ respectively. We retain all individual features as though they all load onto a single factor, each feature is also found to have a statistically significant impact on the performance variables in isolation (i.e., p$<$0.001). \textit{Number of entities} and \textit{number of flows} exceed and perform on par with $F$ respectively; additionally, the overall standard deviation in $R^2$ over all features and $F$ is ~0.04  for both summated dependent variables. This allows us to weight each feature as an individual contributor, as follows:
\vspace{-1em}

\begin{figure}[t]
    \centering
     \begin{subfigure}[b]{0.85\columnwidth}
         \centering
         \includegraphics[width=\textwidth]{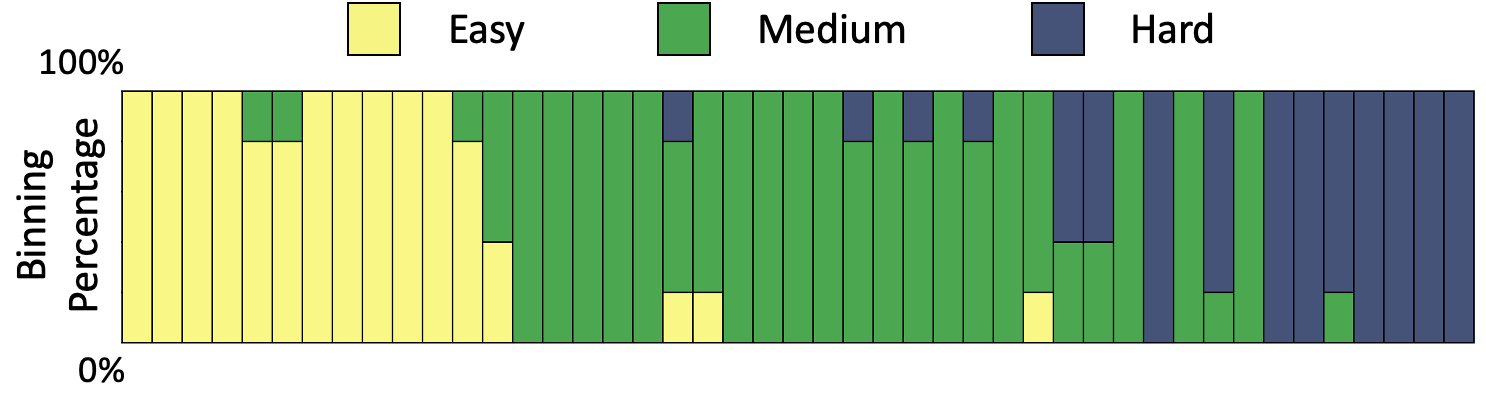}
         \vspace{-2em}
         \caption{$B_3$, Accuracy=$0.788 \pm 0.043$, RMSE=$0.118 \pm 0.017$}
         \label{fig:y equals x}
     \end{subfigure}
     \hfill
     \begin{subfigure}[b]{0.85\columnwidth}
         \centering
         \includegraphics[width=\textwidth]{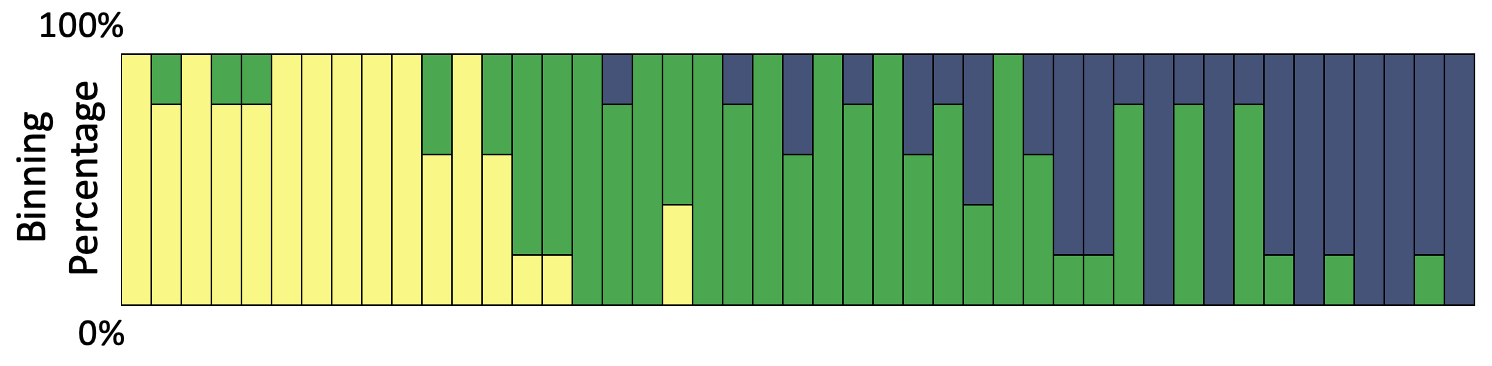}
         \vspace{-2em}
         \caption{$B_4$, Accuracy=$0.742 \pm 0.036$, RMSE=$0.102 \pm 0.019$}
         \label{fig:three sin x}
     \end{subfigure}
     \hfill
     \begin{subfigure}[b]{0.85\columnwidth}
         \centering
         \includegraphics[width=\textwidth]{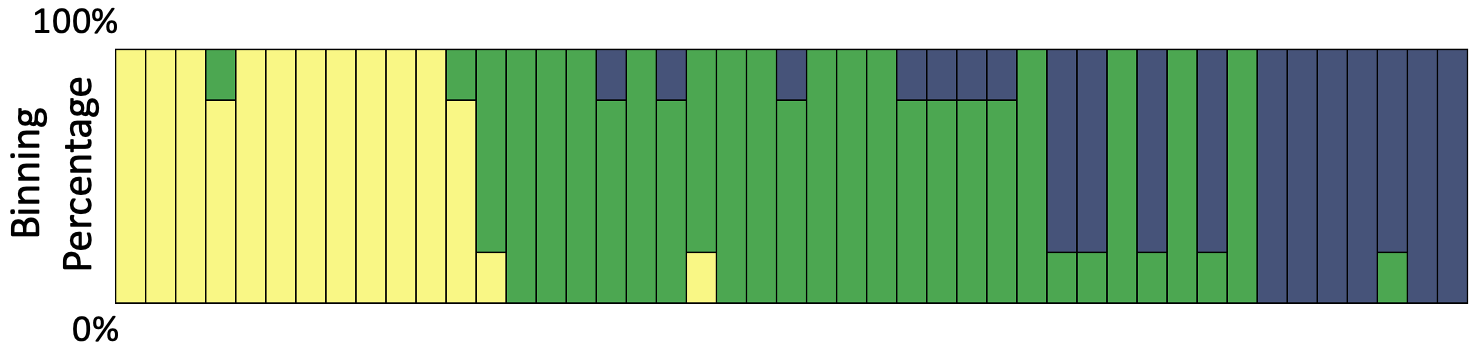}
         \vspace{-2em}
         \caption{$B_{v.c.}$, Accuracy=$0.806 \pm 0.069$, RMSE=$0.257 \pm 0.123$}
         \label{fig:five over x}
     \end{subfigure}
     \hfill
        \caption{These mosaic charts show the diagram-wise Bayesian prediction frequency (i.e., binning percentage) from the(a) $Acc_4$, (b) $Acc_3$, and (c) $S_{v.c.}$ models when classifying our dataset of alluvial diagrams into classes of easy ($<0.33$), medium ($>=0.33 and <0.67$), and hard ($>=0.67$). Each vertical slot corresponds to an alluvial diagram, ordered by $S_a$ value from least to most complex.}
        \label{bayesian}
\end{figure}

\begin{align}
Acc_3(a) = 0.222 (\pm 0.018) t_a + 0.282 (\pm 0.011) e_a \\\nonumber
 + 0.267 (\pm 0.009) f_a + 0.228 (\pm 0.019) c_a
    \label{eq:2}
\end{align}
\vspace{-2em}
\begin{align}
Acc_4(a) = 0.2566 (\pm 0.086) t_a + 0.234 (\pm 0.046) e_a \\\nonumber
+ 0.206 (\pm 0.088) f_a + 0.302 (\pm 0.137) c_a
    \label{eq:3}
\end{align}

We use these equations to construct two Bayesian priors, and train two Bayesian models ($B_{3}$, $B_{4}$) to predict the performance accuracy class of the held-out test-set (20\%), as \textit{easy}/\textit{medium}/\textit{hard} based on chart complexity falling in the lowest, middle, or highest-third bins of the dataset. Prediction trends for $B_3$ and $B_4$ are summarized in Figure \ref{bayesian}(a)-(b), with repeated k-fold cross validation (k=5, n=10), such that each chart is classified at least 5 times. We analyze this chart after the Study~\#2 model ($B_{v.c.}$) is introduced shortly.


\textbf{Study 2 Modelling.}
An initial regression on \textit{perceived complexity} (\%) shows that \textit{number of entities} again serves as the best fitting model ($R^2$=0.637); all independent variables are positively correlated with \textit{perceived complexity}, with p$<$0.001 for all regressions. We also run a regression of $F$ against the \textit{perceived complexity}, and find that \textit{number of entities} still serves as the best predictor. Overall, the features contribute in a more balanced manner compared to the regression results against summated performances in Study~\#1, with relatively closer $R^2$ values ($\sigma=0.03$). 
 
We again update Equation \ref{eq:1}, only now for Study~\#2 we model perceived \textit{visual complexity} ($S_{v.c.}$):
\vspace{-1em}

\begin{align}
S_{v.c.}(a) = 0.240 (\pm 0.036) t_a + 0.247 (\pm 0.061) e_a \\\nonumber
 + 0.314 (\pm 0.073) f_a + 0.197 (\pm 0.129) c_a
    \label{eq:4}
\end{align}

In contrast to Equations~2 and~3, \textit{number of flows} is now the highest weighted factor, despite \textit{number of groups} being the best predictor. We use Equation~4 to construct a Bayesian prior and train a Bayesian model ($B_{v.c.}$) to similarly predict the complexity class of the alluvial diagrams in the test set as \textit{easy/medium/hard} in Figure \ref{bayesian}(c). Compared to $B_{3}$ and $B_{4}$, we find that $B_{v.c.}$ most closely fits complexity classification patterns, and tends to consistently bin diagrams. $B_{3}$ and $B_{4}$ display greater discrepancies during the classification of medium and hard complexity diagrams. $B_{4}$ displays the highest classification variability over epochs; this is expected as task T1 behaves as an outlier, which is taken into account when modelling based on summated accuracy over all tasks ($Acc_4$).


\section{Discussion}

Based on Equations 2--4, we see that the four considered visual features have similar weights in modelling the complexity of an alluvial diagram. Interestingly, we find that when excluding  the ``easy task'' T1 in Equation~2, \textit{number of entities} becomes the most important factor, though when T1 is included, \textit{number of flow crossings} has the highest weight (Equation~3). When considering perceived complexity (Equation~4), \textit{number of flows} is the most important variable. In one sense, the high dependencies between variables makes sense: the \textit{number of entities} in a dataset should highly correlate with the \textit{number of flows} and \textit{number of flow crossings}. However, it is interesting that, depending on how you model complexity, different features seem to emerge as most influential, though this requires further experiments to confirm. It is possible that, as flow arcs are the visual feature in alluvial diagrams that take up the most pixel space (thus leading to higher data-ink ratio~\cite{tufte2001visual}), the \textit{number of flows} has a higher influence on how a person perceives a chart's complexity, albeit only slightly. This also provides a possible reason that \textit{number of flows} is considered the most important feature for the summative variable in the factor analysis conducted in Section~5.

Our approach for modelling complexity (or effectiveness, as defined in Section~2) is extensible to other types of visualizations, including similar techniques like Sankey diagrams, node-link diagrams, and even parallel coordinate plots. However, since Sankey diagrams and node-link diagrams have flows/edges that can go in any direction, more nuanced visual features should be considered, such as flow/edge length and direction. Moreover, any visualization technique (even if dissimilar to alluvial diagrams) can be modelled by selecting an appropriate set of visual features. For example, a bar chart's features could include bar height, bar width, number of bars, the distance between bars, etc. 

We also note some limitations in our studies. For example, Study~\#1 only considers four types of tasks and only measures performance using accuracy. This could be, for example, why T1 had higher performance relative to T2--T4, and therefore did not significantly load during factor analysis. Including results from other metrics such as response time may align T1 more with T2--T4 when modelling complexity. Additionally, generating more diverse alluvial diagrams (i.e., more dataset variation between charts), or considering other visual features (such as entities/flows with varying color hues) will likely affect the resultant models.

\section{Conclusion}

For alluvial diagrams, our research indicates that complexity is, in part, contingent on the user's current task, and that different visual features contribute to complexity differently depending on the current task. However, further experiments are necessary to better understand how alluvial diagram complexity is due to the combined effects of its visual features. Additionally, we believe that our analysis and modelling approach can be employed in evaluating complexity across more diverse visualization techniques, which we plan to study in future experiments.



\acknowledgments{
This research was supported by the U.S. National Science Foundation through grant OAC-1934766.
}

\bibliographystyle{abbrv-doi}

\bibliography{template}

\end{document}